\documentclass[12pt]{article}
\usepackage{epsfig}
\usepackage{color}
\usepackage{amssymb,amsmath}
\usepackage{graphicx}

\newcommand{\bea}{\begin{eqnarray}}
\newcommand{\eea}{\end{eqnarray}}

 \makeatletter
\def\alt{\mathrel{\mathpalette\gl@align<}}
\def\agt{\mathrel{\mathpalette\gl@align>}}
\def\gl@align#1#2{\lower.6ex\vbox{\baselineskip\z@skip\lineskip\z@
\ialign{$\m@th#1\hfil##\hfil$\crcr#2\crcr\sim\crcr}}} \makeatother

\begin{document}
\begin{flushright}
OU-HET 753/2012 \\
UT-12-20
\end{flushright}
\vspace*{1.0cm}

\begin{center}
\baselineskip 20pt 
{\Large\bf 
Minimal Flavor Violation 
in the Minimal $U(1)_{B-L}$ Model
\\
and Resonant Leptogenesis
}
\vspace{1cm}

{\large 
Nobuchika Okada$^{a}$, Yuta Orikasa$^{b}$, Toshifumi Yamada$^{c}$
} \vspace{.5cm}

{\baselineskip 20pt \it
$^{a}$ Department of Physics and Astronomy, University of Alabama, \\
Tuscaloosa, Alabama 35487, USA \\

$^{b}$ Department of Physics, Osaka University, \\
Toyonaka, Osaka 560-0043, Japan \\

$^{c}$ Department of Physics, University of Tokyo,  \\
7-3-1 Bunkyoku, Tokyo 113-0033, Japan}

\vspace{.5cm}

\vspace{1.5cm} {\bf Abstract}
\end{center}

\ \ \ We investigate the resonant leptogenesis scenario
 in the minimally $U(1)_{B-L}$ extended standard model with minimal flavor violation.
In our model, the $U(1)_{B-L}$ gauge symmetry is broken at the TeV scale and
 standard model singlet neutrinos gain Majorana masses of order TeV.
In addition, we introduce a flavor symmetry on the singlet neutrinos at a scale higher than TeV.
The flavor symmetry is explicitly broken by the neutrino Dirac Yukawa coupling,
 which induces splittings in the singlet neutrino Majorana masses at lower scales
 through renormalization group evolutions.
We call this setup ``minimal flavor violation".
The mass-splittings are proportional to the tiny Dirac Yukawa coupling,
 and hence they automatically enhance the CP asymmetry parameter 
 necessary for the resonant leptogenesis mechanism.
In this paper, we calculate the baryon number yield by solving the Boltzmann equations,
 including the effects of $U(1)_{B-L}$ gauge boson that also has TeV scale mass
 and causes washing-out of the singlet neutrinos in the course of thermal leptogenesis.
The Dirac Yukawa coupling for neutrinos
 is fixed in terms of neutrino oscillation data
 and an arbitrary $3 \times 3$ complex-valued orthogonal matrix.
We show that the right amount of baryon number asymmetry 
 can be achieved through thermal leptogenesis in the context of the minimal flavor violation
 with singlet neutrinos and $U(1)_{B-L}$ gauge boson at the TeV scale.
These particles can be discovered at the LHC in the near future.

\thispagestyle{empty}

\newpage

\addtocounter{page}{-1}
\setcounter{footnote}{0}
\baselineskip 18pt
%
\section{Introduction}

\ \ \ The baryon number asymmetry of the Universe has been measured to be \cite{wmap}
\begin{eqnarray}
\frac{n_{B}}{s} &=& ( 0.88 \pm 0.02 ) \ \times \ 10^{-10} \ .
\end{eqnarray}
Since the electroweak baryogenesis scenario based on the standard model (SM) is incapable
 of explaining this amount, we need an extension of SM.
Leptogenesis \cite{leptogenesis} is an appealing scenario,
 as it explains the tiny neutrino masses and the baryon number asymmetry
 simultaneously with heavy singlet neutrinos \cite{seesaw}.
What is problematic in the leptogenesis scenario
 is that it usually cannot be tested experimentally;
the efficiency of leptogenesis is proportional to CP violation which appears in
 interference terms between tree and one-loop diagrams involving singlet neutrinos.
Therefore sufficiently large neutrino Yukawa coupling is required in order that
 the efficiency be high enough to produce the observed amount of baryon asymmetry.
This in turn implies that the singlet neutrinos must be heavy enough to explain the tiny
 neutrino masses, and hence they are beyond the energy scale of collider experiments.

However, if singlet neutrinos have nearly degenerate Majorana masses,
 resonance in loop diagrams enhances the efficiency of leptogenesis,
 making it possible to produce the right amount of baryon asymmetry even with small neutrino
 Yukawa couplings,
 which is called ``resonant leptogenesis" \cite{resonant}.
With the resonant leptogenesis mechanism,
 models with TeV scale singlet neutrinos can explain both the baryon number asymmetry and
 the light neutrino masses.
Still, collider experiments cannot test such models because the singlet neutrinos couple to SM particles
 only through the small Yukawa coupling, and hence they cannot be produced at colliders.
If there is a new gauge interaction that is broken at TeV scale
 and the singlet neutrinos are charged under it,
 we may be able to produce the TeV scale singlet neutrinos at colliders and even measure the amount of CP violation
 through their decays \cite{collider}.
Another problem of models with TeV scale singlet neutrinos
 is its naturalness.
Since successful resonant leptogenesis requires highly degenerate Majorana masses,
 we have to explain the origin of the degeneracy from some new symmetry or mechanism.

In this paper, we propose a model that solves these problems
 and show that it can explain both the light neutrino masses and the baryon number asymmetry
 without losing its experimental accessibility and without requiring unnatural fine-tuning.
In addition to the SM content, 
 our model has $U(1)_{B-L}$ gauge symmetry and contains three singlet neutrinos which are necessary for
 its anomaly cancellation.
This gauge symmetry is broken at the TeV scale so that $U(1)_{B-L}$ gauge boson and 
 the singlet neutrinos obtain masses of order TeV \cite{tev b-l}.
We further introduce a flavor symmetry on the singlet neutrinos at an energy scale above TeV,
 which is broken explicitly by the neutrino Dirac Yukawa coupling.

Because of the flavor symmetry of the singlet neutrinos,
 the differences in their Majorana masses arise solely from 
 the neutrino Dirac Yukawa coupling through renormalization group (RG) evolutions,
 which we call ``minimal flavor violation" for singlet neutrinos.
The ratios of the mass differences over the mass scale
 are therefore of the same order as the neutrino Dirac Yukawa coupling.
It turns out that such mass difference is of an appropriate order to
 produce the right amount of the baryon number asymmetry with resonant enhancement,
 and no fine-tuning is required in our setup.
Moreover, since the differences in the Majorana masses are related to the neutrino Dirac Yukawa coupling,
 our model is more restrictive than the original resonant leptogenesis scenario, where
 the mass difference is a free parameter.

Two of the authors have discussed in ref.~\cite{b-l}
 resonant leptogenesis in the presence of $U(1)_{B-L}$ gauge symmetry
 that is broken at TeV scale.
In that paper, the tiny difference in the Majorana masses of 
 two singlet neutrinos is put in by hand,
 whereas in our paper it arises from RG evolutions involving 
 the neutrino Dirac Yukawa coupling.

Resonant leptogenesis with a flavor symmetry of singlet neutrinos at a high scale
 has been first proposed in ref.~\cite{flavor symmetry}.
However the models in that paper does not contain $U(1)_{B-L}$ gauge symmetry that
 is broken at TeV scale
 because its motivation lies in lowering the reheating temprature of the Universe
 rather than observing singlet neutrinos by collider experiments.
We expect that the washing-out of singlet neutrinos via $U(1)_{B-L}$ gauge interaction
 in the course of thermal leptogenesis reduces the resultant
 baryon number asymmetry, making it difficult to produce the right amount of 
 baryon asymmetry.
This is particularly so when the mass of $U(1)_{B-L}$ gauge boson is larger than twice
 the Majorana masses, because in this case on-shell $U(1)_{B-L}$ gauge boson 
 can mediate the interactions between singlet neutrinos and SM particles. 
From the point of view of collider experiments, however,
 this case is favorable
 because the singlet neutrinos can be produced 
 by the decay of $U(1)_{B-L}$ gauge boson.
We will see that even with the $U(1)_{B-L}$ gauge interaction at the TeV scale,
 we can obtain enough baryon asymmetry through
 resonant leptogenesis with a flavor symmetry of singlet neutrinos at a high scale.

Another feature of our study is that we adopt Casas-Ibarra parametrization \cite{CI}
 for the neutrino Dirac Yukawa coupling.
When the charged lepton Yukawa coupling is neglected,
 the processes of leptogenesis depend only on $(Y_{\nu}^{\dagger})_{ai}(Y_{\nu})_{ib}$,
 where $Y_{\nu}$ denotes the neutrino Dirac Yukawa coupling, $i$ the flavor index of
 lepton doublets and $a,b$ that of singlet neutrinos.
In Casas-Ibarra parametrization,
 this quantity is parametrized in terms of the light neutrino mass eigenvalues and
 an arbitrary complex-valued \textit{orthogonal} matrix,
 independently of the neutrino flavor-mixing matrix.
This parametrization thus makes it clear whether the right amount of baryon number asymmetry
 can be produced in a given setup.

In Section 2, we describe the model which contains $U(1)_{B-L}$ gauge symmetry
 and a flavor symmetry of singlet neutrinos.
In Section 3, we discuss its RG evolutions, which are essential in our model
 as they give rise to the small mass-splittings of Majorana masses necessary 
 for resonant leptogenesis.
In Section 4, we review on the Boltzmann equations that calculate 
 the baryon number yield from leptogenesis.
In Section 5, we calculate the baryon number yield
 for various values of singlet neutrino masses and $U(1)_{B-L}$ gauge boson mass,
 and study whether the observed amount of the baryon asymmetry can be produced.
Section 6 is devoted to the conclusions and discussions.
\\
\\

\section{Setup}

\ \ \ We consider the minimally $U(1)_{B-L}$ extended standard model, where
 the gauge symmetry is $U(1)_{Y} \times U(1)_{B-L} \times SU(2)_{L} \times SU(3)_{c}$
 and right-handed singlet neutrinos, $N_{a} \ (a=1,2,3)$, and a scalar field, $\Phi$, are added to
 the SM matter content.
$\Phi$ obtains VEV and breaks the $U(1)_{B-L}$ gauge symmetry through the Higgs mechanism,
 giving rise to the $U(1)_{B-L}$ gauge boson mass 
 and the Majorana mass term for the singlet neutrinos.
We denote this VEV by $v_{B-L}$.
We are especially interested in the case where $v_{B-L}$ is at the TeV scale,
 and hence the $U(1)_{B-L}$ gauge boson and the singlet neutrinos are observable at collider experiments.
In this case, the components of $Y_{\nu}$ are smaller than about $10^{-5}$.
The field content is summarized in Table 1.
\begin{table}
\begin{center}
\begin{tabular}{|c|c|c|c|} \hline
Field
& $SU(2)_{L}$
& $U(1)_{Y}$
& $U(1)_{B-L}$
\\ \hline
$L_{i}$     & 2 & -1/2 & -1 \\ \hline
$N_{a}$     & 1 & 0            & -1  \\ \hline
$H$         & 2 & -1/2 & 0   \\ \hline
$\Phi$      & 1 & 0            & +2  \\ \hline
\end{tabular}
\end{center}
\caption{Field content in the minimally $U(1)_{B-L}$ extended standard model.
}
\end{table}
We assume that the physical scalar particle that originates from $\Phi$
 after the $U(1)_{B-L}$ symmetry breaking is heavier than the singlet neutrinos,
 and we may neglect it in the discussion of thermal leptogenesis.

We take the combination of the $U(1)_{Y}$ and $U(1)_{B-L}$ gauge bosons
 such that the mixing of their kinetic terms are zero throughout the renormalization group (RG) evolutions.
Then the convariant derivative, $D_{\mu}$, can be written as
\begin{eqnarray}
D_{\mu} \ = \ \partial_{\mu} \ - \ i \sqrt{\frac{3}{5}} g_{1}Y B_{\mu} \ - \ i(\tilde{g}Y+g_{B-L}Y_{B-L}) B_{\mu}^{\prime}
\ - \ ig_{2} \frac{\sigma^{\alpha}}{2} W^{\alpha}_{\mu} \ - \ ig_{3} T^{\alpha} G^{\alpha}_{\mu} \ ,
\end{eqnarray}
 where the coupling $\tilde{g}Y$ reflects the mixing of the $U(1)_{Y}$ and $U(1)_{B-L}$ gauge bosons.
$\tilde{g}, g_{B-L}$ are defined
 so that we have $\tilde{g}=0$ at the $U(1)_{B-L}$ breaking scale.
Since $\Phi$ has $B-L$ charge $2$, its VEV induces the mass of the $U(1)_{B-L}$ gauge boson as
\begin{eqnarray}
M_{Z^{\prime}} &=& 2 \sqrt{2} g_{B-L} v_{B-L} \ .
\end{eqnarray}

The Dirac and Majorana Yukawa couplings for the neutrinos are given by
\begin{eqnarray}
{\cal L} & \supset &
- ( Y_{\nu} )_{i a} \ H L_{i}^{\dagger} N_{a} \ - \ {\rm h.c.} \ - \ \frac{1}{2} (Y_{M})_{a b} \ \Phi N_{a} N_{b} \ - \ {\rm h.c.} 
\end{eqnarray}
Note that $Y_{M}$ is a symmetric matrix with complex values.
The term $\frac{1}{2} (Y_{M})_{a b} \ \Phi N_{a} N_{b}$ becomes the Majorana mass term,
\begin{eqnarray}
\frac{1}{2} v_{B-L} (Y_{M})_{a b} \ N_{a} N_{b} \ ,
\end{eqnarray}
 when $\Phi$ obtains VEV.

The light neutrino mass matrix takes the form:
\begin{eqnarray}
(m_{\nu})_{ij} &=& (Y_{\nu})_{ia} \ (Y_{M}^{-1})_{ab} \ (Y_{\nu}^{T})_{bj} \ \frac{v^{2}}{v_{B-L}}
\end{eqnarray}
We move to the flavor basis where the charged leptons
 and the singlet neutrinos are in their mass eigenstates.
On this basis, we have
\begin{eqnarray}
m^{\dagger}_{\nu} m_{\nu} &=&
U_{MNS} \ \left(
\begin{array}{ccc}
m_{1}^{2} & 0 & 0 \\
0 & m_{2}^{2} & 0 \\
0 & 0 & m_{3}^{2}
\end{array}
\right) \ U_{MNS}^{\dagger} \ ,
\\
Y_{M} &=&
\left(
\begin{array}{ccc}
y_{1} & 0 & 0 \\
0 & y_{2} & 0 \\
0 & 0 & y_{3}
\end{array}
\right) \ ,
\end{eqnarray}
 where $(m_{i})^{2}$'s denote the eigenvalues of the light neutrino mass squared,
 $U_{MNS}$ denotes the neutrino flavor-mixing matrix
 and $y_{i}$'s denote the eigenvalues of $Y_{M}$.
We see from (6), (7) and (8) that $Y_{\nu}$ on this basis can generally be written as
\begin{eqnarray}
(Y_{\nu})_{ia} &=&
U_{MNS}^{*}
\left(
\begin{array}{ccc}
(m_{1}^{2})^{1/4} & 0 & 0 \\
0 & (m_{2}^{2})^{1/4} & 0 \\
0 & 0 & (m_{3}^{2})^{1/4}
\end{array}
\right) \ O \ 
\left(
\begin{array}{ccc}
\sqrt{y_{1}} & 0 & 0 \\
0 & \sqrt{y_{2}} & 0 \\
0 & 0 & \sqrt{y_{3}}
\end{array}
\right) \
\frac{\sqrt{v_{B-L}}}{v} \ ,
\end{eqnarray}
 where $O$ denotes an arbitrary \textit{orthogonal} matrix with \textit{complex} values,
 which appears in Casas-Ibarra parametrization \cite{CI}.
We hereafter call $O$ ``Yukawa kernel matrix".
\\

Let us introduce a symmetry of two generations of singlet neutrinos
 above an energy scale $\Lambda$, which is assumed higher than the TeV scale.
Then we may write
\begin{eqnarray}
Y_{M}[\Lambda] &\propto& 
\left(
\begin{array}{ccc}
1 & 0 & 0 \\
0 & 1 & 0 \\
0 & 0 & r
\end{array}
\right) \ 
\end{eqnarray}
 in the basis where $Y_{M}$ is diagonal.
This is the key assumption that realizes 
 the resonant leptogenesis without fine-tuning of the parameters.
Below the scale $\Lambda$, $Y_{\nu}$ explicitly breaks the flavor symmetry 
 and comes into effect through RG evolutions.
The difference between the first two components of $Y_{M}$ at lower scales
 is proportional to $Y_{\nu}^{\dagger} Y_{\nu}$,
 and hence the singlet neutrinos obtain Majorana masses
 whose difference between the first two components
 is proportional to $Y_{\nu}^{\dagger} Y_{\nu}$
 when $\Phi$ gains VEV.
This mass difference is essential for successful resonant leptogenesis.

Our flavor symmetry is inspired by a class of $A_4$ symmetric models where
 the three flavors of the singlet neutrinos transform as $3$ representation of $A_4$ group
 (see \textit{e.g.} Table 1 of ref.~\cite{a4} for a list of such models),
 although we do not specify the details of the flavor symmetry.
If the singlet neutrinos are in $3$ representation and the $B-L$ breaking field $\Phi$ is in $1$ representation
 of $A_4$ group,
 the Majorana Yukawa coupling will be common for the three flavors.
In realistic $A_4$ models, 
 especially when one takes into account the observed value of the neutrino mixing angle $\theta_{13}$ 
 \cite{theta13},
 $A_4$ symmetry should be spontaneously broken below some scale.
In our model, the spontaneous breaking of a flavor symmetry is 
 effectively described by the Dirac Yukawa coupling that explicitly violates the flavor symmetry
 and that comes into play below the scale $\Lambda$.
\\
\\

\section{RG Evolutions}

\ \ \ In the following discussion, we set the $U(1)_{B-L}$ breaking scale, $M_{0}$,
 at $\vert y \vert v_{B-L}$, where $y$ is of the order of the eigenvalues of $Y_{M}$.
The results do not change with $O(1)$ variation of the $U(1)_{B-L}$ breaking scale.
We further make an approximation
 that the $U(1)_{B-L}$ gauge boson mass, $M_{Z^{\prime}} = 2 \sqrt{2} g_{B-L} v_{B-L}$,
 is of the same order as $M_{0}$.

The RG equations for the U(1) gauge couplings, $g_{1}, g_{B-L}, \tilde{g}$, are 
 given as follows, where $M_{0}$ is the $U(1)_{B-L}$ breaking scale:
\begin{eqnarray}
16 \pi^{2} \mu \frac{{\rm d}}{{\rm d} \mu} g_{1} &=& \frac{53}{15} g_{1}^{3} \ \ \ {\rm for} \ \mu < m_{t} \ , \nonumber
\\ 
16 \pi^{2} \mu \frac{{\rm d}}{{\rm d} \mu} g_{1} &=& \frac{41}{10} g_{1}^{3} \ \ \ {\rm for} \ \mu > m_{t} \ , 
\\ 16 \pi^{2} \mu \frac{{\rm d}}{{\rm d} \mu} g_{B-L} &=& 12 g_{B-L}^{3} 
\ + \ 2 \cdot \frac{16}{3} g_{B-L}^{2} \tilde{g} \ + \ \frac{41}{6} g_{B-L} \tilde{g}^{2}  \ \ \ {\rm for} \ \mu > M_{0} \ , 
\\ 16 \pi^{2} \mu \frac{{\rm d}}{{\rm d} \mu} \tilde{g} &=& \frac{41}{6} \tilde{g} (\tilde{g}^{2} + \frac{6}{5} g_{1}^{2})
\ + \ 2 \cdot \frac{16}{3} g_{B-L} (\tilde{g}^{2} + \frac{3}{5} g_{1}^{2})
\ + \ 12 g_{B-L}^{2} \tilde{g} \ \ \ {\rm for} \ \mu > M_{0} . \nonumber \\
\end{eqnarray}
The RG equations for $g_{2}, g_{3}$ are the same as in the standard model.

The RG equations for the Yukawa couplings $Y_{M}, Y_{\nu}^{\dagger} Y_{\nu}$ and the top Yukawa coupling, $y_{t}$, are given as :
\begin{eqnarray}
16 \pi^{2} \mu \frac{{\rm d}}{{\rm d} \mu} (Y_{M})_{a b} &=& (Y_{\nu}^{\dagger} Y_{\nu})_{a c} (Y_{M})_{c b} \ + \  (Y_{M})_{a c} (Y_{\nu}^{\dagger} Y_{\nu})_{c b} \nonumber
\\ &+& (Y_{M} Y_{M}^{*} Y_{M})_{a b} \ + \ \frac{1}{2} (Y_{M})_{a b} \ {\rm tr}[ Y_{M} Y_{M}^{\dagger} ]
\ - \ 6 (g_{1}^{\prime})^{2} (Y_{M})_{a b} \ \ \ {\rm for} \ \mu > M_{0} \ ,
\nonumber \\
\\ 16 \pi^{2} \mu \frac{{\rm d}}{{\rm d} \mu} (Y_{\nu}^{\dagger} Y_{\nu})_{a b} &=& (Y_{\nu}^{\dagger} Y_{\nu})_{a b} 
\ ( \ 6 y_{t}^{2} \ - \ \frac{9}{10} g_{1}^{2} \ - \ \frac{9}{2} g_{2}^{2} 
\ - \ 12 g_{B-L}^{2} \ - \ \frac{3}{2} \tilde{g}^{2} \ - \ 6 g_{B-L} \tilde{g} \ ) \nonumber
\\ &+& \frac{1}{2} ( Y_{M} Y_{M}^{*} Y_{\nu}^{\dagger} Y_{\nu} )_{a b} \ 
+ \ \frac{1}{2} ( Y_{\nu}^{\dagger} Y_{\nu} Y_{M} Y_{M}^{*} )_{a b} \ \ \ {\rm for} \ \mu > M_{0} \ ,
\\ 16 \pi^{2} \mu \frac{{\rm d}}{{\rm d} \mu} y_{t} &=& \frac{9}{2} y_{t}^{3} \ - \ y_{t} \ 
( \ \frac{17}{20} g_{1}^{2} \ + \ \frac{9}{4} g_{2}^{2} \ + \ 8 g_{3}^{2} \ ) \ \ \ {\rm for} \ m_{t} < \mu < M_{0} \ , \nonumber
\\ 16 \pi^{2} \mu \frac{{\rm d}}{{\rm d} \mu} y_{t} &=& \frac{9}{2} y_{t}^{3} \ - \ y_{t} \
( \ \frac{17}{20} g_{1}^{2} + \frac{9}{4} g_{2}^{2} + 8 g_{3}^{2} 
+ \frac{17}{12} \tilde{g}^{2} + \frac{2}{3} g_{B-L}^{2} + \frac{5}{3} \tilde{g} g_{B-L} \ )
 \ \ \ {\rm for} \ \mu > M_{0} . \nonumber \\
\end{eqnarray}
In the RG equations above, we neglect the Yukawa couplings except the top Yukawa coupling.
Note that, when the charged lepton Yukawa couplings are neglected,
 the RG evolution for $Y_{\nu}$ can be expressed in terms of that for $Y_{\nu}^{\dagger}Y_{\nu}$.

The boundary conditions for the RG equations are
\begin{eqnarray}
Y_{M}[\Lambda] &=&
\left(
\begin{array}{ccc}
\tilde{y} & 0 & 0 \\
0 & \tilde{y} & 0 \\
0 & 0 & \tilde{y_{3}}
\end{array}
\right) \ ,
\\ 
Y_{\nu}^{\dagger}Y_{\nu}[M_{0}] &\simeq& 
\left(
\begin{array}{ccc}
\sqrt{ y^{*} } & 0 & 0 \\
0 & \sqrt{ y^{*} } & 0 \\
0 & 0 & \sqrt{ y_{3}^{*} }
\end{array}
\right) \
O^{\dagger} \
\left(
\begin{array}{ccc}
\sqrt{ m_{1}^{2} } & 0 & 0 \\
0 & \sqrt{  m_{2}^{2} } & 0 \\
0 & 0 & \sqrt{ m_{3}^{2} }
\end{array}
\right) \
O \
\left(
\begin{array}{ccc}
\sqrt{ y } & 0 & 0 \\
0 & \sqrt{ y } & 0 \\
0 & 0 & \sqrt{ y_{3} }
\end{array}
\right) \
\frac{v_{B-L}}{v^{2}} \ \nonumber \\
\\
\tilde{g}[M_{0}] &=& 0 \ .
\end{eqnarray}
Note that $Y^{\dagger}_{\nu}Y_{\nu}$ is independent of the neutrino flavor-mixing matrix, $U_{MNS}$.
In the first and second equations,
 $\tilde{y}, \tilde{y_{3}}$ and $y, y_{3}$ are related through RG evolutions
 from the scale $\Lambda$ to $M_{0}$.
In order to obtain their relation, 
 we solve eq. (14) neglecting the term $Y_{\nu}^{\dagger}Y_{\nu}$.
\\

We numerically solve the RG equations using the following best fit values (with $3 \sigma$ ranges)
 from the Particle Data Group \cite{pdg}, which were derived
 assuming non-zero $\theta_{13}$:
\begin{eqnarray}
m_{2}^{2} - m_{1}^{2} &=& 7.58 \ \ ^{+0.60}_{-0.59} \ \times \ 10^{-5} \ {\rm eV}^{2} \ , 
\\
\vert m_{3}^{2} - m_{1}^{2} \vert &=& 2.35 \ ^{+ 0.32}_{- 0.29} \ \times \ 10^{-3} \ {\rm eV}^{2} \ , 
\end{eqnarray}
Our analysis is done for both the normal hierarchy ($m_{3}^{2} > m_{1}^{2}$)
 and the inverted hierarchy ($m_{3}^{2} < m_{1}^{2}$)
 of the light neutrino masses.
We fix the mass of the lightest neutrino at zero, i.e.,
 we set $m_{1}^{2}=0$ for the normal hierarchy case
 and $m_{3}^{2}=0$ for the inverted hierarchy case.
The Yukawa kernel matrix $O$ can be any 3$\times$3 orthogonal matrix.
However, since the purpose of this paper is to show that
 successful leptogenesis can be realized in our framework,
 we here restrict our analysis to a specific choice of the matrix $O$
 that can be expressed as
\begin{eqnarray}
O &=& 
\left(
\begin{array}{ccc}
0 & 0 & 1 \\
\cos \alpha & \sin \alpha & 0 \\
-\sin \alpha & \cos \alpha & 0
\end{array}
\right) \ {\rm for \ the \ normal \ hierarchy \ case,} \nonumber
\\
O &=& 
\left(
\begin{array}{ccc}
\cos \alpha & \sin \alpha & 0 \\
-\sin \alpha & \cos \alpha & 0 \\
0 & 0 & 1
\end{array}
\right) \ {\rm for \ the \ inverted \ hierarchy \ case,}
\end{eqnarray}
 where $\alpha$ takes a \textit{complex} value.
With these assumptions, the upper-left 2$\times$2 part of
 $Y_{\nu}^{\dagger}Y_{\nu}[M_{0}]$
 takes the following simple form for the normal hierarchy case:
\begin{eqnarray}
Y_{\nu}^{\dagger}Y_{\nu}[M_{0}] \vert_{2\times 2} &\simeq&
\left(
\begin{array}{cc}
\cos \alpha^{*} & -\sin \alpha^{*} \\
\sin \alpha^{*} & \cos \alpha^{*} \\
\end{array}
\right)
\left(
\begin{array}{cc}
\sqrt{m_{2}^{2}} & 0 \\
0 & \sqrt{m_{3}^{2}} \\
\end{array}
\right)
\left(
\begin{array}{cc}
\cos \alpha & \sin \alpha \\
-\sin \alpha & \cos \alpha \\
\end{array}
\right)
\frac{ \vert y \vert v_{B-L} }{v^{2}} \ .
\end{eqnarray}
For the inverted hierarchy case, $\sqrt{m_{2}^{2}}$ and $\sqrt{m_{3}^{2}}$ in the above formula
 are replaced with $\sqrt{m_{1}^{2}}$ and $\sqrt{m_{2}^{2}}$.

We calculate the mass spectrum for various values of $\alpha$
 and substitute it into the Boltzmann equations that calculate the baryon number yield.
Since we are assuming a symmetry on only two generations of singlet neutrinos,
 it is natural that the other singlet neutrino has distinctively different Majorana mass.
This singlet neutrino then does not contribute to resonant leptogenesis.
We here neglect this singlet neutrino and 
 make a two-flavor approximation, i.e.,
 we only consider the two generations of singlet neutrinos,
 on solving the Boltzmann equations.
\\
\\

\section{Baryon Number Asymmetry from Leptogenesis}

\ \ \ At high temperatures,
 the singlet neutrinos as well as the SM particles
 are in thermal equilibrium.
(Remember that the $U(1)_{B-L}$ gauge boson mediates interactions between
 the singlet neutrinos and other particles.)
As the Universe cools down and the temperature becomes below 
 the order of $M_{0}$,
 the singlet neutrinos decouple from the thermal bath
 and decay into the Higgs doublet and the SM leptons.
Because of the CP phases of $Y_{\nu}$, the decays of the singlet neutrinos
 exhibit a CP asymmetry, which gives rise to non-zero lepton number density,
 or equivalently non-zero $B-L$ charge density.
The lepton number density is converted into the baryon number asymmetry through
 the sphaleron process \cite{sph1, sph2} when 
 the temperature of the Universe is higher than
 the decoupling temprature of the sphaleron process $T_{sph} \sim 150$ GeV \cite{Tsph}.

To estimate the resultant baryon number asymmetry, we solve the Boltzmann equations
 governing the number densities of the particles concerned \cite{KT}.
We normalize the temperature, $T$, of the radiation-dominated Universe as
\begin{eqnarray*}
z &\equiv& y v_{B-L} \ / \ T \ ,
\end{eqnarray*}
 where $y v_{B-L}$ approximately equals to
 the mass of the two singlet neutrinos that are relevant to leptogenesis.
We respectively denote the yields, $n/s$ ($n$: number density of a particle species, $s$: entropy density),
 of the singlet neutrinos $N_{a}$ ($a=1,2$) and the $B-L$ charge by
 $Y_{N_{a}}$, $Y_{B-L}$.

\begin{figure}
  \begin{center}
   \includegraphics[width=120mm]{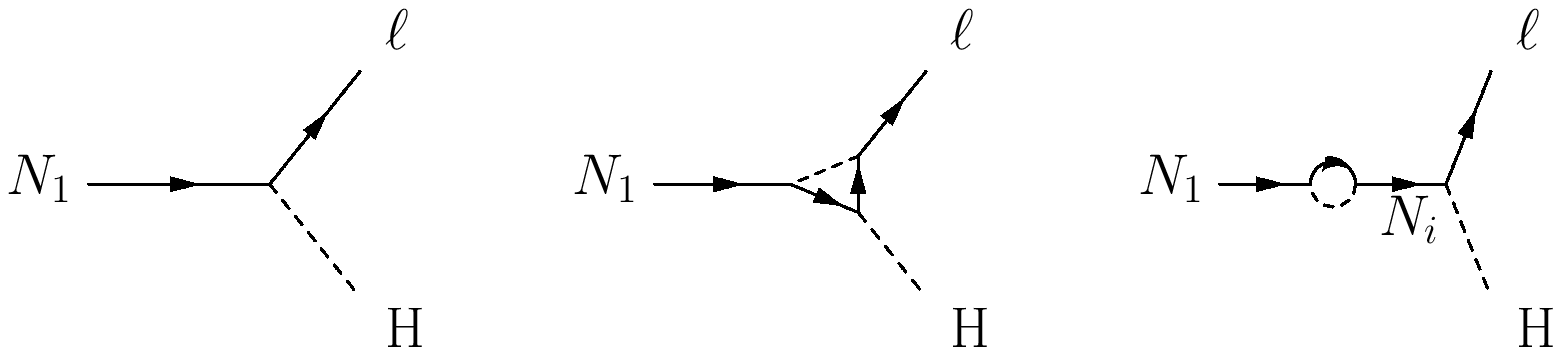}
  \end{center}
 \caption{Decay of a singlet neutrino at tree and one-loop levels.
 }
\end{figure}
The CP asymmetry parameter $\epsilon_{a}$ for the decay of singlet neutrino $N_{a}$
 is defined as
\begin{eqnarray}
\epsilon_{a} &\equiv& 
\frac{ \sum_{j} [ \Gamma( N_{a} \rightarrow L_{j} H^{\dagger} )
- \Gamma( N_{a} \rightarrow L_{j}^{\dagger} H ) ] }{
\sum_{j} [ \Gamma( N_{a} \rightarrow L_{j} H^{\dagger} )
+ \Gamma( N_{a} \rightarrow L_{j}^{\dagger} H ) ] } \ .
\end{eqnarray}
Non-zero $\epsilon_{a}$ comes from the interference of the diagrams in Figure 1.
It is generally given by
\begin{eqnarray}
\epsilon_{a} &=& -\sum_{b \neq a} \frac{M_{a}}{M_{b}} 
\frac{\Gamma_{b}}{M_{b}} 
\left( \frac{V_{b}}{2} + S_{b} \right)
\frac{ {\rm Im} \left[ (Y_{\nu}^{\dagger}Y_{\nu})_{ab}^{2} \right] }
{ (Y_{\nu}^{\dagger}Y_{\nu})_{aa} (Y_{\nu}^{\dagger}Y_{\nu})_{bb} } \ ,
\end{eqnarray}
 where $M_{a}$, $\Gamma_{a}$ denote the Majorana mass and the decay width
 of the singlet neutrino $N_{a}$ and we have
 $\Gamma_{a} = M_{a} (Y_{\nu}^{\dagger}Y_{\nu})_{aa} /  8 \pi$.
$V_{b}$ reflects the interference of the left and the middle diagrams
 in Figure 1 and $S_{b}$ that of the left and the right diagrams.
They are given by
\begin{eqnarray}
V_{b} &=& 2 \frac{M_{b}^{2}}{M_{a}^{2}} \left[ \left(
1+\frac{M_{b}^{2}}{M_{a}^{2}} \right) \log \left(
1+\frac{M_{a}^{2}}{M_{b}^{2}} \right) - 1 \right] \ ,
\\
S_{b} &=& \frac{ M_{b}^{2} (M_{b}^{2} - M_{a}^{2}) }
{ (M_{b}^{2} - M_{a}^{2})^{2} + M_{a}^{2} \Gamma_{b}^{2} } \ .
\end{eqnarray}
In our model, the mass difference of the relevant two singlet neutrinos,
 $M_{2}^{2} - M_{1}^{2}$, comes from the RG evolutions
 from the scale $\Lambda$ down to $M_{0}$.
Therefore it is estimated as
\begin{eqnarray}
M_{2}^{2} - M_{1}^{2} &\propto& Y_{\nu}^{\dagger} Y_{\nu} \ (y v_{B-L})^2 \ .
\end{eqnarray}
Since we are assuming $y v_{B-L} \sim $TeV, 
 it follows from eq. (9) that the components of $Y_{\nu}$
 is at least as small as $( \Delta m_{atm}^2 /  {\rm TeV} )^{1/4} \sim 10^{-4}$.
Then $S_{1}, S_{2}$ are enhanced by the factor of $(1/Y_{\nu}^{\dagger} Y_{\nu}) \sim 10^{8}$,
 whereas $V_{1}, V_{2}$ remain $O(1)$.
In eq. (25), this enhancement cancels the suppression factor $Y_{\nu}^{\dagger} Y_{\nu}$
 that comes from the term $\Gamma_{b}/M_{b}$,
 and we have $O(1)$ values of $\epsilon_{1}, \epsilon_{2}$.
This is the heart of resonant leptogenesis naturally realized with the minimal flavor violation.

The Boltzmann equations that govern the yields of the singlet neutrinos $N_{1}, N_{2}$
 and the $B-L$ charge are written as
\begin{eqnarray}
\frac{{\rm d} Y_{N_{1}}}{{\rm d} z} &=& - \frac{z}{s H(y v_{B-L})} 
\left[ \left( \frac{ Y_{N_{1}} }{ Y^{eq}_{N_{1}} } - 1 \right) \gamma_{D_{1}}
+ \left( \left( \frac{ Y_{N_{1}} }{ Y^{eq}_{N_{1}} } \right)^{2} - 1 \right) \gamma_{Z^{\prime}} \right] \ ,
\\
\frac{{\rm d} Y_{N_{2}}}{{\rm d} z} &=& - \frac{z}{s H(y v_{B-L})} 
\left[ \left( \frac{ Y_{N_{2}} }{ Y^{eq}_{N_{2}} } - 1 \right) \gamma_{D_{2}}
+ \left( \left( \frac{ Y_{N_{2}} }{ Y^{eq}_{N_{2}} } \right)^{2} - 1 \right) \gamma_{Z^{\prime}} \right] \ ,
\\
\frac{{\rm d} Y_{B-L}}{{\rm d} z} &=& - \frac{z}{s H(y v_{B-L})} 
\left[ \sum_{a=1}^{2} 
\left( \frac{1}{2} \frac{ Y_{B-L} }{ Y^{eq}_{L} } 
- \epsilon_{a} \left( \frac{ Y_{N_{a}} }{ Y^{eq}_{N_{a}} } - 1 \right) \right) \gamma_{D_{a}} \right] \ .
\end{eqnarray}
$Y^{eq}_{N_{a}}, Y^{eq}_{L}$ respectively denote
 the yields of the singlet neutrino $N_{a}$ and the lepton doublets $L_{i}$s,
 and $H(y v_{B-L})$ denotes the Hubble constant at the temprature $T=y v_{B-L}$.
$\gamma_{D_{a}}$ is the space-time density of the decay events of $N_{a}$,
 and $\gamma_{Z^{\prime}}$ the space-time density of the processes 
 mediated by the $U(1)_{B-L}$ gauge boson:
 $N_{a} N_{a} \leftrightarrow f \bar{f}$, where $f$ indicates a standard model fermion.
We neglect other terms that give negligibly small contributions to the yields,
 in accordance with ref. \cite{b-l}.
The general form of the Boltzmann equation and expressions of $\gamma$'s
 are given in ref. \cite{boltz}.

At small $z$, $U(1)_{B-L}$ gauge interaction keeps the singlet neutrinos in thermal equilibrium.
Below the sphaleron decoupling temperature, the sphaleron process ends and
 the $B-L$ charge density is no longer converted to
 the baryon number density.
Noting these,
 we numerically solved the Boltzmann equations from $z=0.001$ 
 to $z=y v_{B-L}/T_{sph}$
 with the initial conditions:
\begin{eqnarray}
Y_{N_{a}}(z=0.001) &=& Y^{eq}_{N_{a}} (z=0.001) \ ,
\\
Y_{B-L}(z=0.001) &=& 0 \ ,
\end{eqnarray}
 where we simply assumed that there is initially no $B-L$ charge density.
The resultant $B-L$ charge density is converted to 
 the baryon number density through the sphaleron process \cite{sph2}.
The yield of the baryon number, $n_{B}/s$, is related to $Y_{B-L}$
 by the following formula:
\begin{eqnarray}
\frac{n_{B}}{s} &=& \frac{28}{79} Y_{B-L} ( z=y v_{B-L}/T_{sph} ) \ .
\end{eqnarray}
\\
\\

\section{Numerical Analysis on Resonant Leptogenesis}

\ \ \ Solving the RG equations in Section 3 and the Boltzmann equations in Section 4,
 we calculate the baryon number density for various values of the parameters,
 and compare it with the observed value eq. (1).
The parameters are 
\\

The $U(1)_{B-L}$ gauge boson mass (at the scale $M_{0}$): \ $M_{Z^{\prime}} = 2 \sqrt{2} g_{B-L} v_{B-L}$

The $U(1)_{B-L}$ gauge coupling constant (at the scale $M_{0}$): \ $g_{B-L}$

Scale above which the flavor symmetry of two singlet neutrinos appears: \ $\Lambda$

Approximate value of the mass of the singlet neutrinos (at the scale $M_{0}$): \ $y v_{B-L}$

The parameter of the Yukawa kernel matrix $O$: \ $\alpha \ \ (\in \mathbb{C})$
\\

We set the $U(1)_{B-L}$ breaking scale $M_{0}$ at $y v_{B-L}$,
 and the sphaleron decoupling temprature $T_{sph}$ at $150$ GeV \cite{Tsph}.
We adopt the following values:
\begin{eqnarray*}
\Lambda &=& 10^{5} \ {\rm GeV} \ ,
\\
\frac{g_{B-L}^{2}}{4 \pi} &=& 0.005 \ .
\end{eqnarray*}
 while substituting various values into $y v_{B-L}$ and $M_{Z^{\prime}}$.
We assume the normal hierarchy of the light neutrino masses.
With these parameters,
 we make contour plots of the baryon number yield on the plane of real and complex parts
 of $\alpha$ that parametrizes the neutrino Dirac Yukawa matrix $Y_{\nu}$.
The baryon number density depends on $\alpha$ only through the form of $Y_{\nu}^{\dagger}Y_{\nu}$, (17).
Since $Y_{\nu}^{\dagger}Y_{\nu}$ is invariant under the transformation: 
 $Re[ \alpha ] \ \rightarrow \ Re[ \alpha ] \ + \ \pi$,
 we restrict the range of $\alpha$ to $0 \leq Re[ \alpha ] \leq \pi$.

We also make contour plots of the baryon number yield
 for $\Lambda = 10^{15}$ GeV, for $\frac{g_{B-L}^{2}}{4 \pi} = 0.02$
 and for the inverted hierarchy of the light neutrino masses,
 to study the impacts of these parameters on the result.
\\

In Figure 2, we show how the CP asymmetry parameters, $\epsilon_{1}, \ \epsilon_{2}$,
 evaluated at the scale $M_{0}$ depend on $\Lambda$.
Here we set $g_{B-L}^{2}/4\pi = 0.005$, $y v_{B-L} = 2$ TeV, $M_{Z^{\prime}} = 5$ TeV and
 $\alpha = 2 + i$.
\begin{figure}[htbp]
  \begin{center}
   \includegraphics[width=120mm]{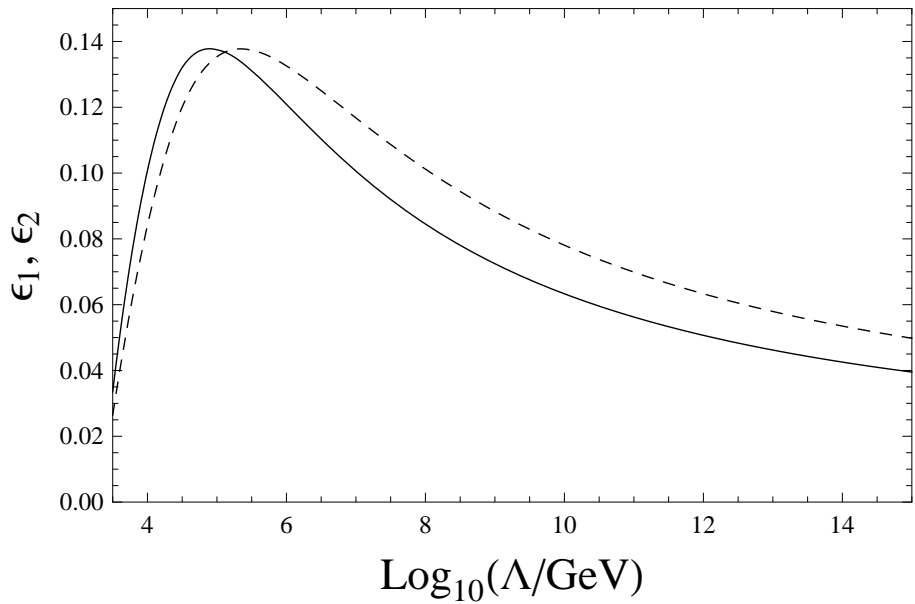}
  \end{center}
 \caption{CP asymmetry parameters, $\epsilon_{1}$ (solid) and $\epsilon_{2}$ (dashed), 
 at the scale $M_{0}$ 
 as functions of the singlet neutrino symmetry scale, $\Lambda$.
 }
\end{figure}
We see that above the scale of $10^{4}$ GeV, the CP asymmetry parameters 
 of order $0.1$ are automatically realized by the RG evolutions.
Also, they only weakly depend on the scale of $\Lambda$,
 which means that the baryon number yield is not affected by
 the scale above which the flavor symmetry of singlet neutrinos appears.
\\

On the left and right of Figure 3, 4, 5 and  6, we show contour plots of the baryon number yield, $n_{B}/s$,
 on the plane of $(Re[\alpha], Im[\alpha])$
 for $(M_{Z^{\prime}}, y v_{B-L}) = (3, 1)$, $(3, 2)$, $(4, 1.5)$, $(4, 3)$, $(5, 2)$, $(5, 3)$ TeV.
Thick solid (red) contours denote the regions where
 the resultant baryon number yield fits with the observed value eq. (1).
Cases with $(M_{Z^{\prime}}, y v_{B-L}) = (3, 1), \ (4, 1.5), \ (5, 2)$ TeV
 are particularly interesting because the singlet neutrinos can be produced
 by the decay of $U(1)_{B-L}$ gauge boson if it is produced as a resonant state
 at colliders.
\begin{figure}[htbp]
 \begin{minipage}{0.5\hsize}
  \begin{center}
   \includegraphics[width=80mm]{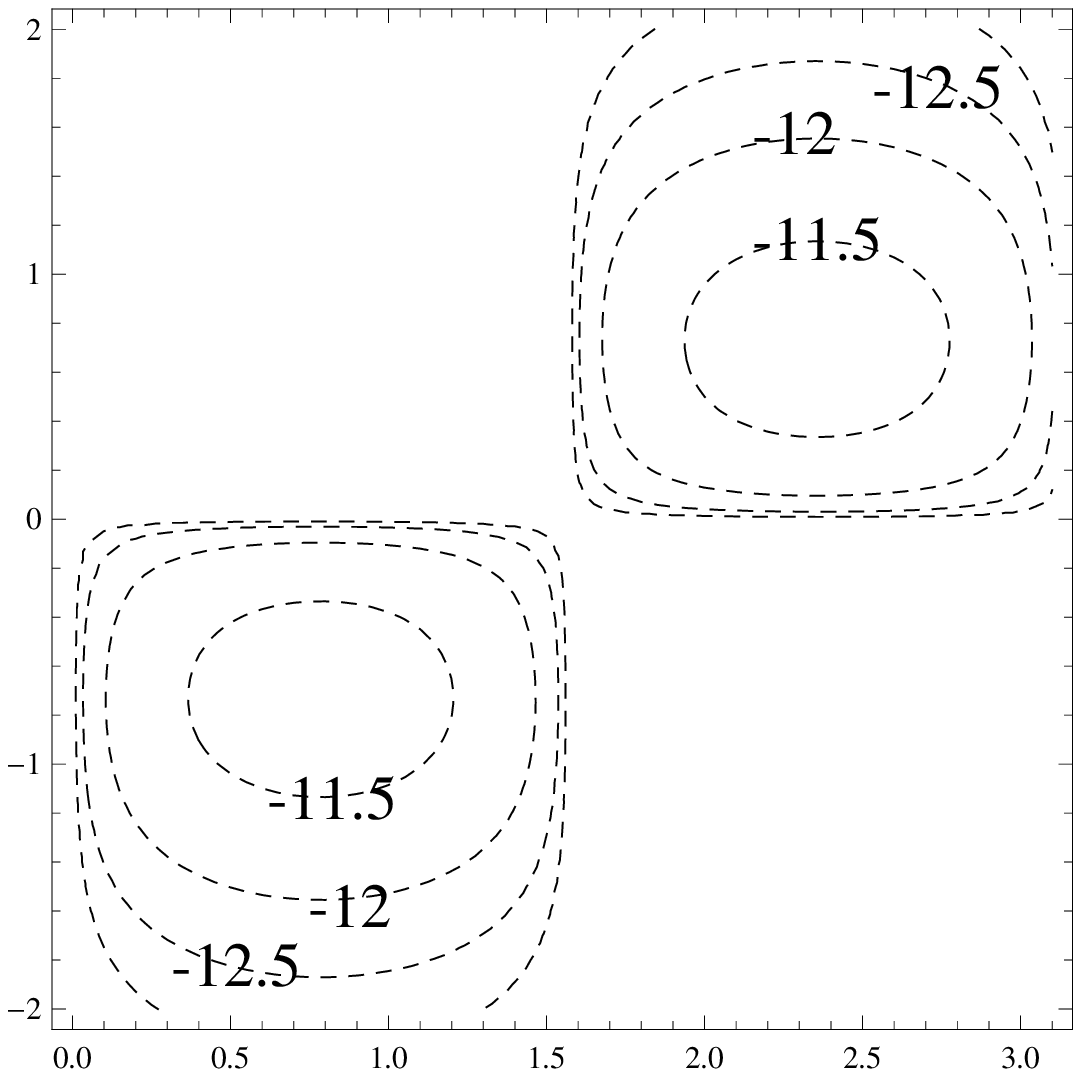}
  \end{center}
 \end{minipage}
 \begin{minipage}{0.5\hsize}
  \begin{center}
   \includegraphics[width=80mm]{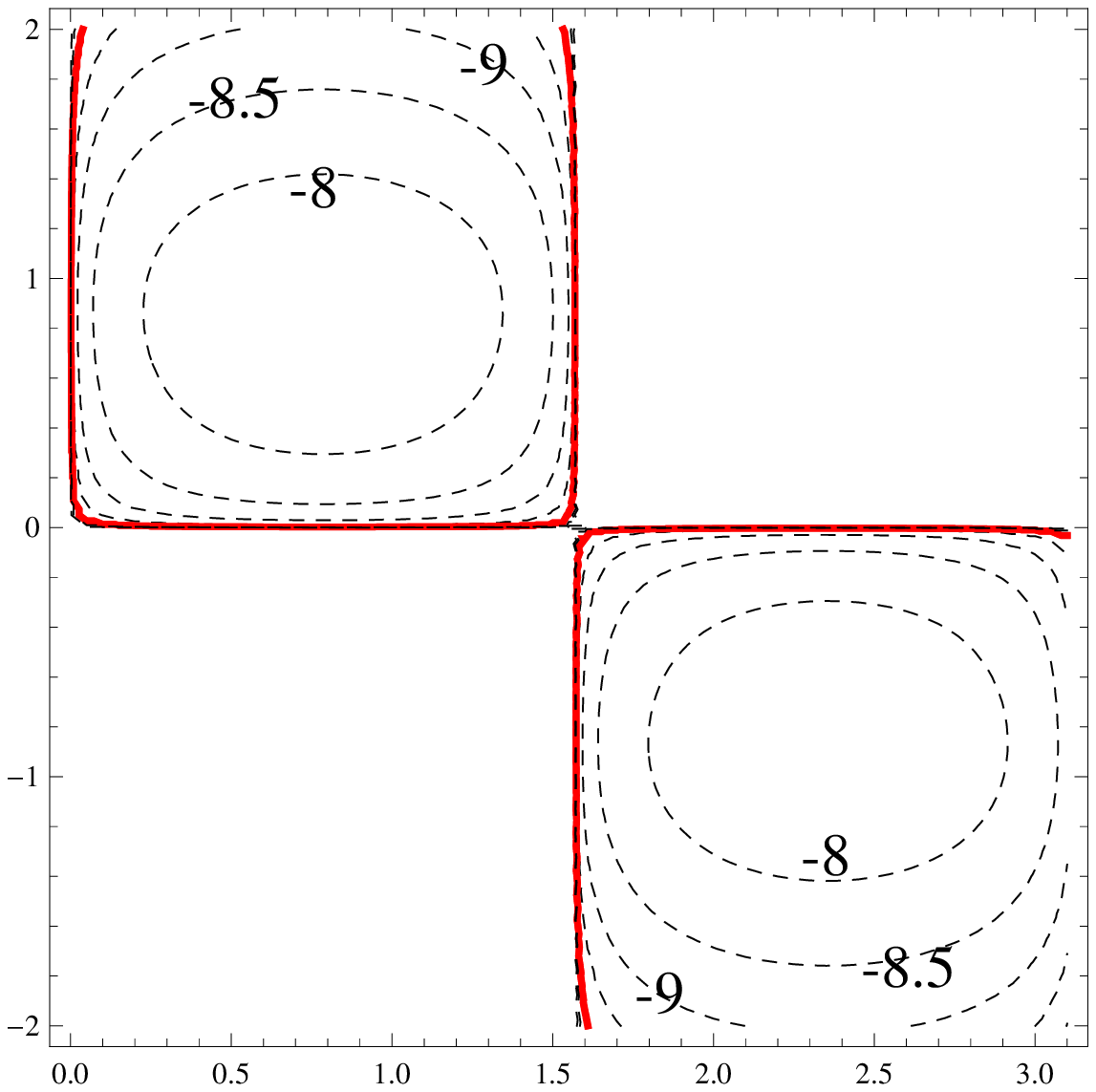}
  \end{center}
 \end{minipage}
 \caption{Contour plots of the baryon number yield, $n_{B}/s$, as a function 
 of the complex parameter $\alpha$ that parametrizes the neutrino Dirac Yukawa matrix $Y_{\nu}$.
 We take $g_{B-L}^{2}/4\pi = 0.005$ and $\Lambda=10^{5}$ GeV, and assume
 the normal hierarchy of the light neutrino masses.
 \textit{Left}: $(M_{Z^{\prime}}, y v_{B-L}) = (3, 1)$ TeV, 
 \textit{Right}: $(M_{Z^{\prime}}, y v_{B-L}) = (3, 2)$ TeV.
 The numbers on the contours denote the corresponding values of $\log_{10} (n_{B}/s)$.
 The thick red line denote the region where the baryon number yield fits 
 with the observed value eq. (1).}
\end{figure}
\begin{figure}[htbp]
 \begin{minipage}{0.5\hsize}
  \begin{center}
   \includegraphics[width=80mm]{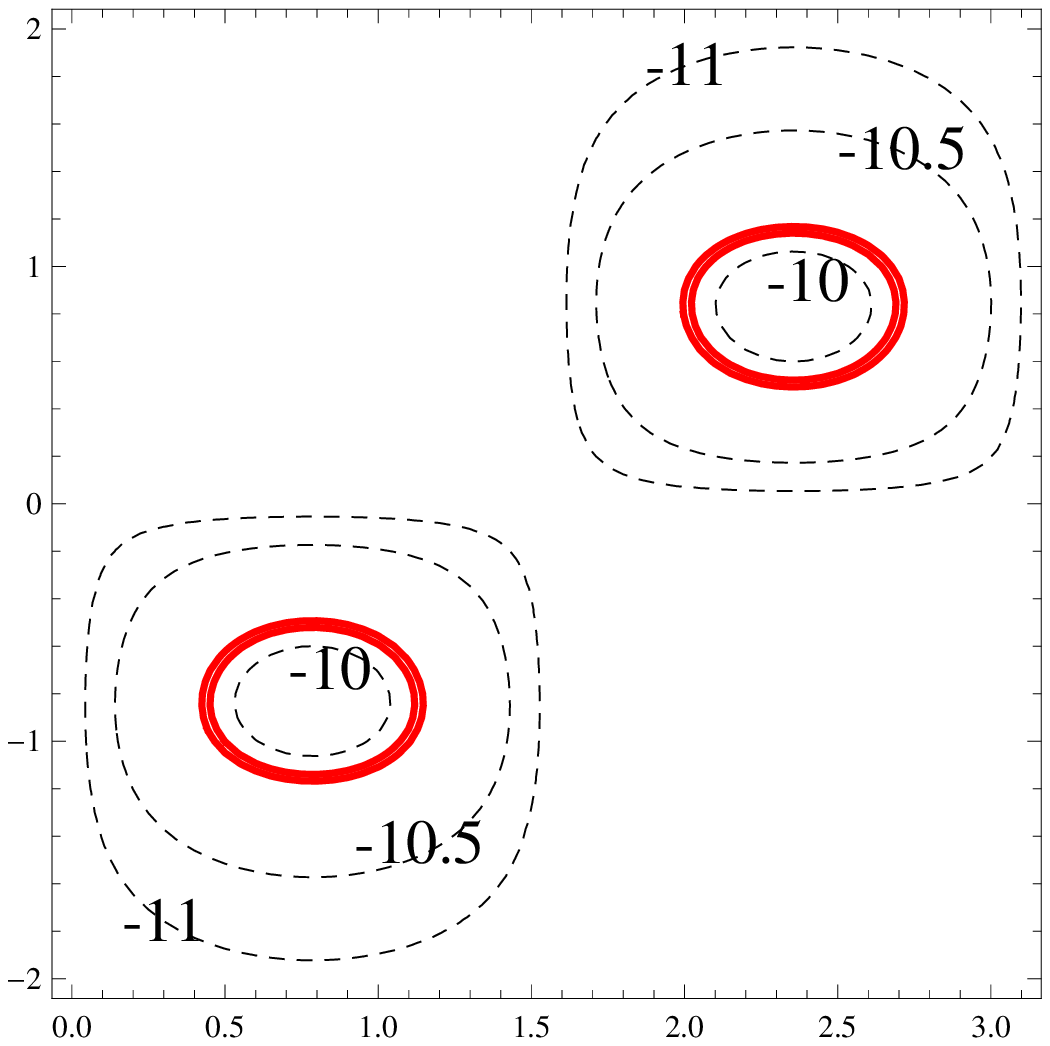}
  \end{center}
 \end{minipage}
 \begin{minipage}{0.5\hsize}
  \begin{center}
   \includegraphics[width=80mm]{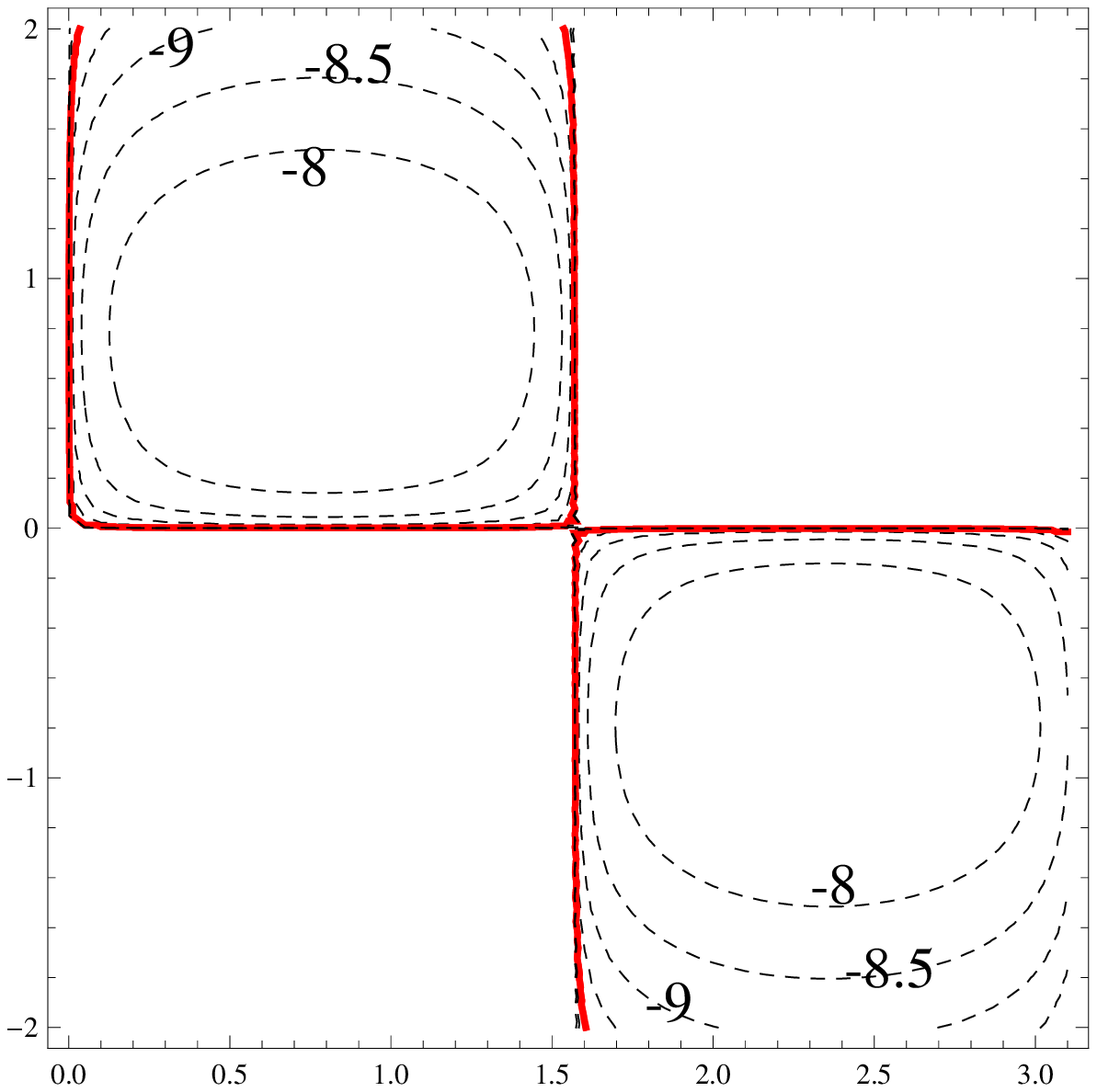}
  \end{center}
 \end{minipage}
 \caption{Same as Figure 3, but we here have
 \textit{Left}: $(M_{Z^{\prime}}, y v_{B-L}) = (4, 1.5)$ TeV, 
 \textit{Right}: $(M_{Z^{\prime}}, y v_{B-L}) = (4, 3)$ TeV.
 }
\end{figure}
\begin{figure}[htbp]
 \begin{minipage}{0.5\hsize}
  \begin{center}
   \includegraphics[width=80mm]{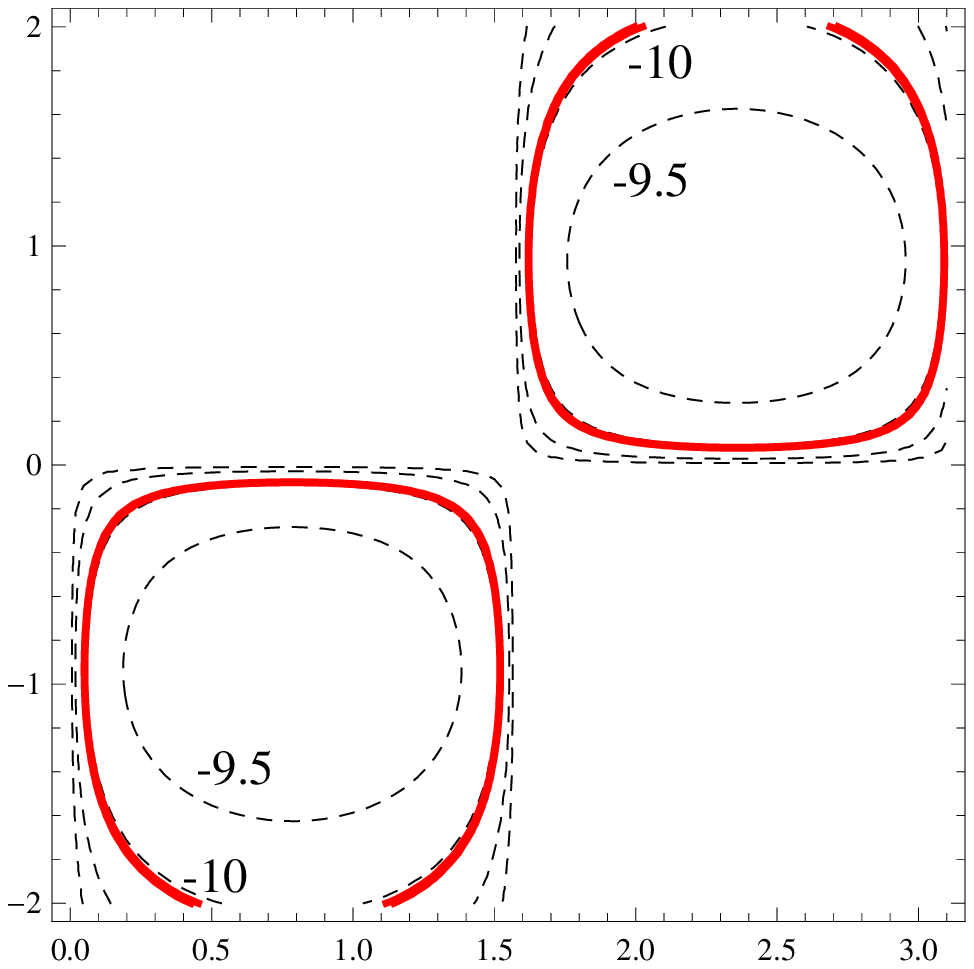}
  \end{center}
 \end{minipage}
 \begin{minipage}{0.5\hsize}
  \begin{center}
   \includegraphics[width=80mm]{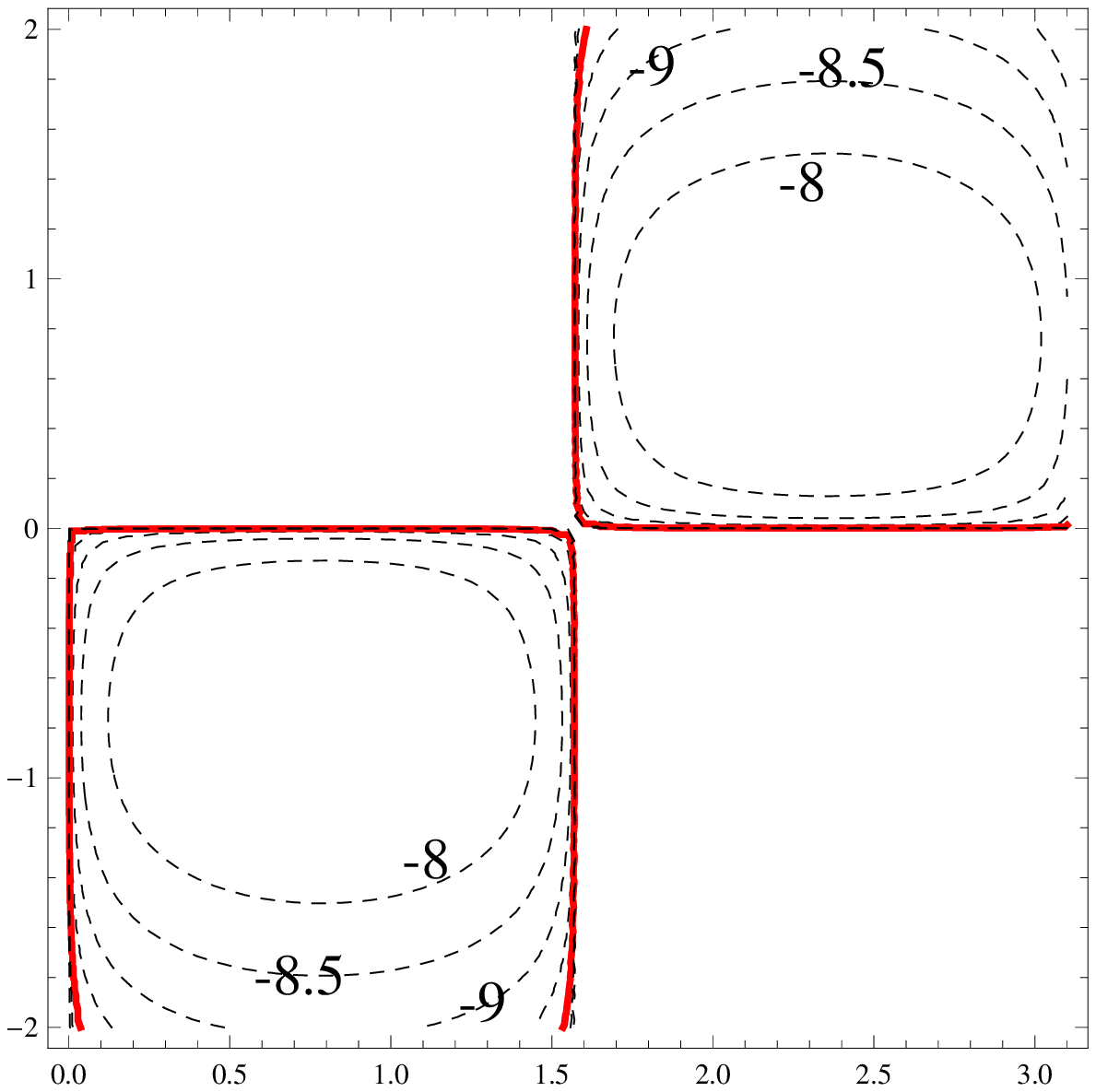}
  \end{center}
 \end{minipage}
 \caption{Same as Figure 3, but we here have
 \textit{Left}: $(M_{Z^{\prime}}, y v_{B-L}) = (5, 2)$ TeV, 
 \textit{Right}: $(M_{Z^{\prime}}, y v_{B-L}) = (5, 3)$ TeV.
 }
\end{figure}
\.
\\

Figure 6 is the same contour plot as the left of Figure 5, 
 but we here take $\Lambda = 10^{15}$ GeV.
Figure 7 is the same contour plot as the left of Figure 5, 
 but we here take $g_{B-L}^{2}/4\pi = 0.02$.
Figure 8 is the same contour plot as the left of Figure 5, 
 but we here assume the inverted hierarchy of the light neutrino masses.
\begin{figure}
  \begin{center}
   \includegraphics[width=80mm]{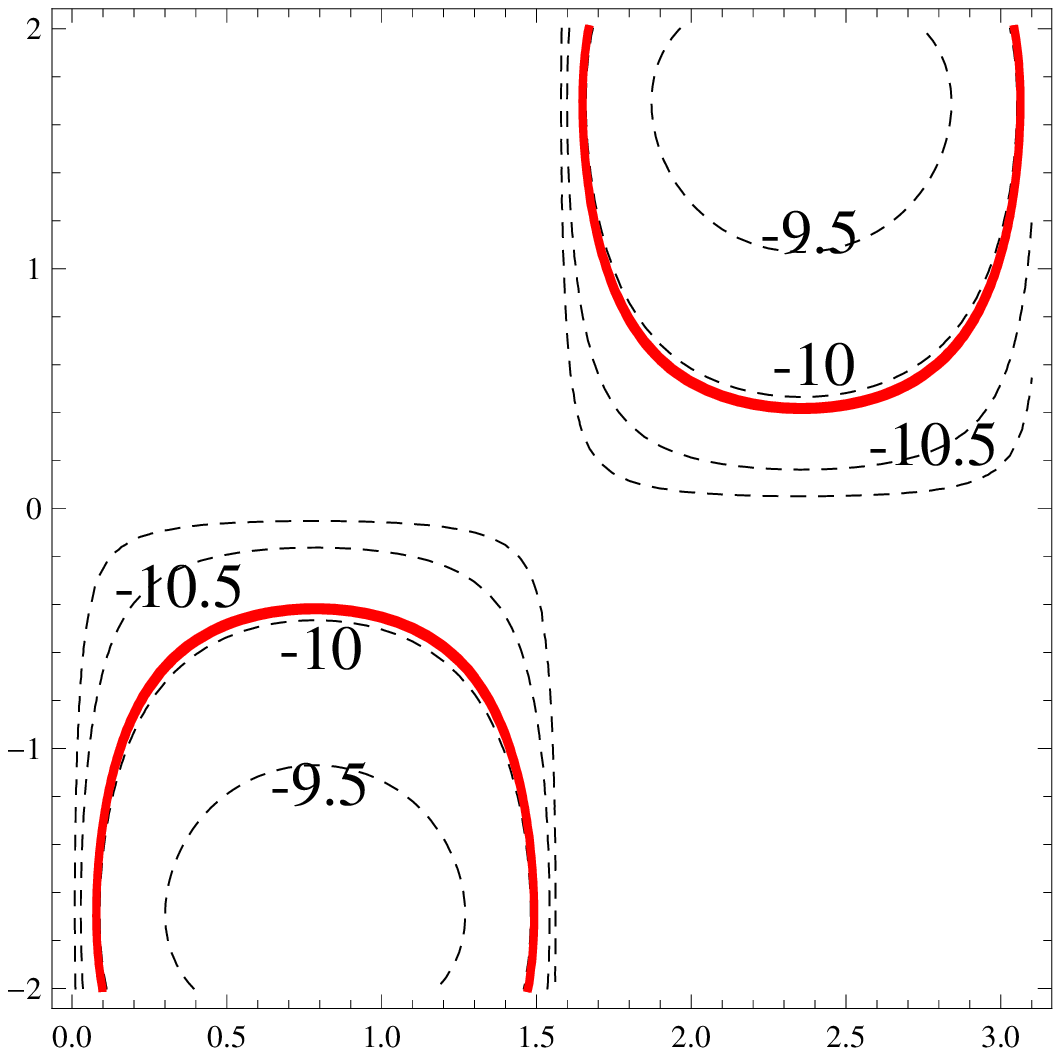}
  \end{center}
 \caption{Same as the left of Figure 5, but we here take $\Lambda = 10^{15}$ GeV.
 }
\end{figure}
\begin{figure}
  \begin{center}
   \includegraphics[width=80mm]{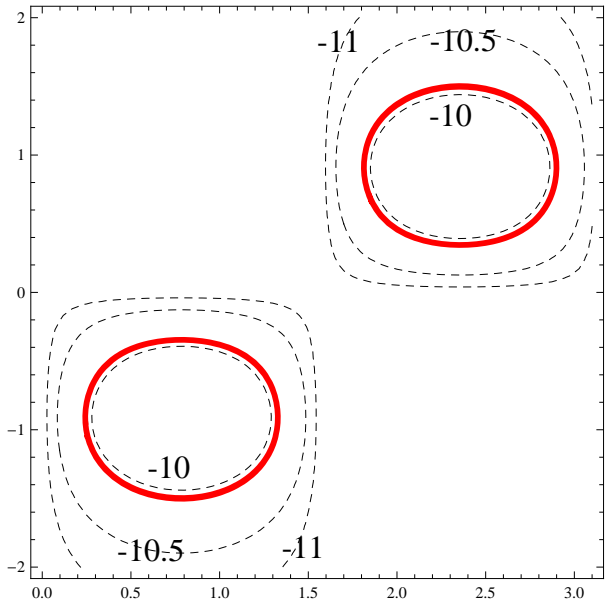}
  \end{center}
 \caption{Same as the left of Figure 5, but we here take
 $g_{B-L}^{2}/4\pi = 0.02$. 
 }
\end{figure}
\begin{figure}
  \begin{center}
   \includegraphics[width=80mm]{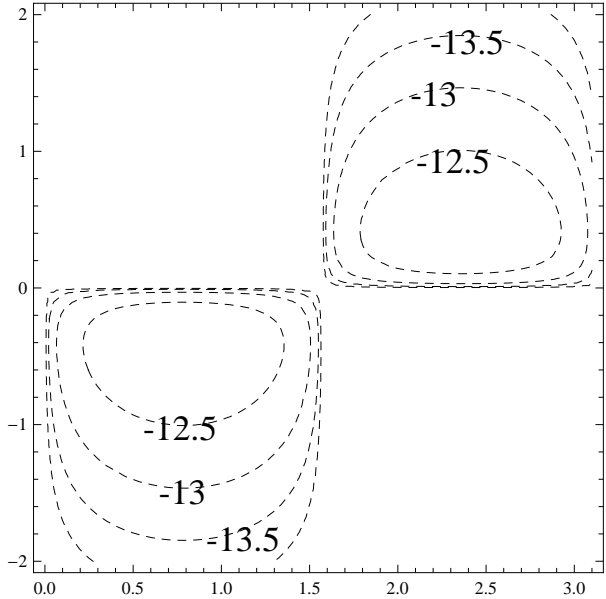}
  \end{center}
 \caption{Same as the left of Figure 5, but we here assume the inverted hierarchy of the light neutrino masses.
 }
\end{figure}
\\

We notice from the left panels of Figure 4 and 5 that
 a sufficiently large amount of the baryon number yield
 can be produced in benchmark points with
 $(M_{Z^{\prime}}, \ y v_{B-L}) = (4, 1.5)$ TeV and $(5, 2)$ TeV,
 and $g_{B-L}^{2}/4 \pi = 0.005$,
 which are of interest for collider studies.
Moreover the observed baryon number yield, eq. (1), is obtained for moderate values of $(Re[\alpha], Im[\alpha])$,
 which implies that no fine-tuning is required for the neutrino Dirac Yukawa coupling
 in order to reproduce the observed value.

From Figures 3, 4 and 5,
 we see that the heavier the singlet neutrinos are,
 the more baryon number yield we have.
This is simply due to the fact that 
 leptogenesis ends at the sphaleron decoupling temperature, which is assumed
 to be $T_{sph} \simeq 150$ GeV.
If the singlet neutrino mass is smaller, less singlet neutrinos decouple from
 the thermal bath and do out-of-equilibrium decay before the radiation of the Universe cools down to 
 the sphaleron decoupling temperature.
We also notice that we have more baryon number yield when $M_{Z^{\prime}} < 2 y v_{B-L}$
 than otherwise.
This is because in the latter case, an on-shell $U(1)_{B-L}$ gauge boson
 mediates the processes: 
 $N_{a} N_{a} \leftrightarrow Z^{\prime} \leftrightarrow f \bar{f}$
 ($N_{a}$: singlet neutrino, $f$: standard model fermion),
 and singlet neutrinos are washed out more efficiently.

From Figure 5 and 6, we confirm that
 the baryon number yield hardly depends on the scale $\Lambda$, 
 above which the flavor symmetry of the singlet neutrinos appears.
This is in accord with Figure 2, which shows that
 the CP asymmetry parameters vary only by factor 2 with the change of $\Lambda$.

Comparing Figure 7 with Figure 5, 
 we see that the baryon number yield decreases for larger $g_{B-L}$,
 which reflects the fact that the washing-out of singlet neutrinos
 is more efficient when the $U(1)_{B-L}$ gauge coupling is stronger.

Finally from Figure 5 and 8, we see that 
 the baryon number yield drastically decreases with the inverted hierarchy of
 light neutrino masses.
This is because the neutrino Dirac Yukawa coupling squared, eq. (18),
 cannot have large CP-violating phase 
 when two of the mass eigenvalues for the light neutrinos 
 are nearly degenerate,
 as in the inverted hierarchy case.
\\
\\

\section{Conclusions and Discussions}

\ \ \ We have studied the resonant leptogenesis scenario
 in the model where SM is extended with $U(1)_{B-L}$ gauge symmetry 
 which is broken by the Higgs mechanism at the TeV scale and with TeV scale singlet neutrinos.
We further assume a flavor symmetry on two generations of the singlet neutrinos
 which is explicitly broken by the neutrino Dirac Yukawa coupling at a scale above TeV.
Renormalization group evolutions generate a splitting of the singlet neutrino Majorana masses
 that is proportional to the neutrino Dirac Yukawa coupling.
Hence the CP asymmetry parameters for the singlet neutrino decay process
 are automatically enhanced to order 0.1 by the resonance,
 even though the neutrino Dirac Yukawa coupling is much smaller.
With the above setup, we have calculated the baryon number yield by solving the Boltzmann equations,
 taking into account the effects of the $U(1)_{B-L}$ gauge boson that can wash out the singlet neutrinos.
We parametrize the neutrino Dirac Yukawa coupling
 by the light neutrino mass eigenvalues obtained from the neutrino oscillation data 
 and an arbitrary $2 \times 2$ complex-valued orthogonal matrix.
We have derived the baryon number yield for various values of the singlet neutrino mass and
 $U(1)_{B-L}$ gauge boson mass,
 and compared it with the observed value, eq. (1).
We have shown that a sufficiently large amount of baryon number yield is generated
 with several TeV masses for the singlet neutrinos and $U(1)_{B-L}$ gauge boson.
The most notable result is that the right amount of baryon number yield
 can be obtained for $(M_{Z^{\prime}}, \ y v_{B-L}) = (4, 1.5)$ TeV and $(5, 2)$ TeV
 with $g_{B-L}^{2}/4 \pi = 0.005$,
 in spite of the fact that an on-shell $U(1)_{B-L}$ gauge boson washes out
 singlet neutrinos in the course of thermal leptogenesis.
\\

Let us discuss the implications of the above result on collider physics.
At the LHC with $\sqrt{s}=14$ TeV,
 the cross section for a $U(1)_{B-L}$ gauge boson that eventually decays into $\mu^{+} \mu^{-}$
 is given by
\begin{eqnarray}
\sigma( pp \rightarrow Z^{\prime} \rightarrow \mu^{+} \mu^{-} ) &=& 0.200 \ {\rm fb} 
\ \ \ {\rm for} \ M_{Z^{\prime}}=4 \ {\rm TeV} \ {\rm and} \ g_{B-L}^{2}/4 \pi = 0.005, \\
\sigma( pp \rightarrow Z^{\prime} \rightarrow \mu^{+} \mu^{-} ) &=& 0.0271 \ {\rm fb} 
\ \ \ {\rm for} \ M_{Z^{\prime}}=5 \ {\rm TeV} \ {\rm and} \ g_{B-L}^{2}/4 \pi = 0.005,
\end{eqnarray}
 where we have integrated over the muon invariant mass 
 from $M_{Z^{\prime}}-1$ TeV to $M_{Z^{\prime}}+1$ TeV.
SM background is negligibly small in these regions.
We therefore expect to discover the $U(1)_{B-L}$ gauge boson 
 with the integrated luminosity below 100 fb$^{-1}$ in these cases.
The production cross section for the two generations of singlet neutrinos
 is given by $(i=1,2)$
\begin{eqnarray}
\sigma( pp \rightarrow Z^{\prime} \rightarrow N_i \bar{N}_i ) &=& 0.138 \ {\rm fb} 
\ \ \ {\rm for} \ (M_{Z^{\prime}}, \ y v_{B-L}) = (4, 1.5) \ {\rm TeV} \ 
{\rm and} \ g_{B-L}^{2}/4 \pi = 0.005, \nonumber \\
\\
\sigma( pp \rightarrow Z^{\prime} \rightarrow N_i \bar{N}_i ) &=& 0.0181 \ {\rm fb} 
\ \ \ {\rm for} \ (M_{Z^{\prime}}, \ y v_{B-L}) = (5, 2) \ {\rm TeV} \ 
\ {\rm and} \ g_{B-L}^{2}/4 \pi = 0.005. \nonumber \\
\end{eqnarray}
A singlet neutrino decays into a light neutrino and the Higgs boson or the longitudinal component
 of $Z$ boson, or into a charged lepton and the longitudinal component of $W$ boson.
By combining several decay modes, it is possible to identify singlet neutrinos \cite{basso}
 with the integrated luminosity of several hundred fb$^{-1}$
 even in the case of $(M_{Z^{\prime}}, \ y v_{B-L}) = (5, 2)$ TeV.
\\
\\

\section*{Acknowledgements}

\ \ \ This work is supported in part by 
the DOE Grant No. DE-FG02-10ER41714 (N.O.), 
Grant-in-Aid for Scientific research 
on Innovative Areas, No. 23104009 (Y.O.), 
and the grant of the Japan Society for the Promotion
of Science, No. 23-3599 (T.Y.). 
\\
\\

\section*{Appendix}

\ \ \ The number density $n_\psi$ of a particle $\psi$ 
 (a right-handed neutrino in our case) 
 with mass $m_\psi$ in the early universe is evaluated 
 by solving the Boltzmann equation of the form \cite{KT}, 
\begin{eqnarray}
 \frac{dY_\psi}{dz}=-\frac{z}{sH(m_\psi)}\sum_{a,i,j,\cdots}
 \left[\frac{Y_\psi Y_a \cdots}{Y_\psi^{eq} Y_a^{eq} \cdots}
 \gamma^{eq}(\psi+a+\cdots\rightarrow i+j+\cdots)\right.\nonumber\\
 \left.-\frac{Y_i Y_j\cdots}{Y_i^{eq} Y_j^{eq}\cdots}
 \gamma^{eq}(i+j+\cdots\rightarrow\psi+a+\cdots)\right], 
\label{boltzeq}
\end{eqnarray}
where $Y_\psi =n_\psi/s$ is the ratio of $n_\psi$ 
 and the entropy density $s$, $z=\frac{m_\psi}{T}$, 
 and $H(m_\psi)$ is the Hubble parameter at a temperature $T=m_\psi$. 
The right hand side of eq.~(\ref{boltzeq}) describes 
 the interactions that change number of $\psi$,  
 and $\gamma^{eq}$ is the space-time density of scatterings 
 in thermal equilibrium. 
For a dilute gas we take into account 
 decays, two-particle scatterings and the corresponding back reactions. 
One finds, for a decay the particle $\psi$, 
\begin{equation}
 \gamma_D=\gamma^{eq}(\psi\rightarrow i+j+\cdots)
=n_\psi^{eq}\frac{K_1(z)}{K_2(z)}\tilde{\Gamma}_{rs}, 
\end{equation}
where $K_1$ and $K_2$ are the modified Bessel functions,  
 and $ \tilde{\Gamma}_{rs}$ is the decay width. 
For two body scattering one has 
\begin{equation}
 \gamma\left(\psi+a\leftrightarrow i+j+\cdots\right)
 =\frac{T}{64\pi^4}\int^\infty_{(m_\psi+m_a)^2} ds 
 \hat{\sigma}(s)\sqrt{s}K_1\left(\frac{\sqrt{s}}{T}\right), 
\end{equation}
where $s$ is the squared center-of-mass energy and the reduced cross 
 section $\hat{\sigma}$ for the process 
 $\psi+a\leftrightarrow i+j+\cdots$ is related 
 to the usual total cross section $\sigma(s)$ by 
\begin{equation}
 \hat{\sigma}(s)=\frac{8}{s}\left[(p_\psi\cdot p_a)^2
-m_\psi^2 m_a^2\right]\sigma(s). 
\end{equation}

The total reduced cross section for the process 
 $f+\bar{f} \rightarrow N+N$ mediated 
 by the $Z'$ boson ( $f$ denotes the SM fermions) 
 is given by 
\begin{equation}
 \hat{\sigma}_{Z^\prime}(s)=\frac{104\pi}{3}\alpha_{B-L}^2
\frac{\sqrt{x}}{(x-y)^2+yc}\left(x-4\right)^{\frac{3}{2}},  
\end{equation}
 where $y=\frac{m_{Z'}^2}{m_N^2}$, and 
 $c=\left(\tilde{\Gamma}_{Z^\prime}/{M_1}\right)^2$ 
 with the decay width of $Z'$ boson 
\begin{equation}
\tilde{\Gamma}_{Z^\prime}=\frac{\alpha_{B-L} m_{Z^\prime}}{6}
\left[ 3\left(1-\frac{4}{y}\right)^\frac{3}{2}\theta(y-4)
+13\right]. 
\end{equation}
\\
\\

\end{document}